\documentclass[12pt,a4paper]{article}
\usepackage[T2A]{fontenc}
\usepackage[cp1251]{inputenc}
\usepackage[english]{babel}
\usepackage{amsmath,amssymb,amsfonts}
\usepackage{bm,latexsym,euscript,textcomp}
\usepackage{graphicx,color,graphics,caption,epsf,dcolumn}
\usepackage{natbib}
\usepackage{hyperref}
\usepackage{epstopdf}
\usepackage{pdflscape}
\usepackage{rotating}
\usepackage{booktabs}
\usepackage{gensymb}

\newcommand{\beq}{\begin{equation}}
\newcommand{\eeq}{\end{equation}}
\newcommand{\pt}{\partial}

\begin{document}

\title{\Large \bf  Comment on "The Tropospheric Land-Sea Warming Contrast as the Driver of Tropical Sea Level
Pressure Changes" by Bayr and Dommenget}

\author{A. M. Makarieva$^{1,2}$\thanks{\textit{Corresponding author.} {E-mail: ammakarieva@gmail.com}}, V. G. Gorshkov$^{1,2}$, A.V. Nefiodov$^1$,\\ D.  Sheil$^{3,4,5}$, A. D. Nobre$^6$, B.-L. Li$^2$}

\date{\vspace{-5ex}}

\maketitle

\noindent
$^1$Theoretical Physics Division, Petersburg Nuclear Physics Institute, 188300 Gatchina, St. Petersburg, Russia
$^2$XIEG-UCR International Center for Arid Land Ecology, University of California, Riverside 92521-0124, USA
$^3$Norwegian University of Life Sciences, \AA s, Norway
$^4$School of Environment, Science and Engineering, Southern Cross University, PO Box 157, Lismore, NSW 2480, Australia;
$^5$Center for International Forestry Research, PO Box 0113 BOCBD, Bogor 16000, Indonesia;
$^6$Centro de Ci\^{e}ncia do Sistema Terrestre INPE, S\~{a}o Jos\'{e} dos Campos SP 12227-010, Brazil.

\begin{abstract}
\citet{bayr13} proposed a model of temperature-driven air redistribution to quantify the ratio between changes of sea level pressure $p_s$
and mean tropospheric temperature $T_a$ in the tropics. This model assumes that the height of the tropical troposphere is isobaric.
Here problems with this model are identified. A revised relationship between $p_s$ and $T_a$ is derived governed by two parameters --
the isobaric and isothermal heights -- rather than just one. Further insight is provided by the model of \citet{lindzen87} which was the first
to use the concept of isobaric height to relate tropical $p_s$ to air temperature, and did this by assuming that isobaric height is always around
3~km and isothermal height is likewise near constant. Observational data, presented here, show that neither of these heights is spatially universal
nor do their mean values match previous assumptions. Analyses show that the ratio of the long-term changes in $p_s$ and $T_a$ associated with land-sea
temperature contrasts in a warming climate -- the focus of \citet{bayr13} -- is in fact determined by the corresponding ratio of spatial
differences in the annual mean $p_s$ and $T_a$. The latter ratio, reflecting lower pressure at higher temperature in the tropics, is dominated
by meridional pressure and temperature differences rather than by land-sea contrasts.
Considerations of isobaric heights are shown to be unable to predict either spatial or temporal variation in $p_s$.
As noted by \citet{bayr13}, the role of moisture dynamics
in generating sea level pressure variation remains in need of further theoretical investigations.
\end{abstract}

\section{Introduction}

Low-level tropical winds are generally linked to convection, but the physical processes and relationships remain a matter of interest and discussion.
Indeed, our incomplete understanding of the physical principles governing low-level circulation is manifested by the inability of atmospheric models to
replicate the terrestrial water cycle \citep{ma06,hagemann11} as well as by the challenge of confidently predicting precipitation and air
circulation \citep[e.g.,][]{an11,acha12,huang13}.  One recurring question is whether the release of latent heat in the upper atmosphere
generates sufficient moisture convergence in the lower atmosphere to feed convection. The observed relationship between sea level pressure
and surface temperature (with warm areas having low pressure) is regarded as evidence that low-level convergence is, rather, driven by the temperature gradients
\citep[see discussions by][]{lindzen87,neelin89,sobel06,back09,an11}.

The physical rationale behind surface pressure gradients driven by surface temperature gradients is that a gaseous atmosphere
held by a gravitational field cannot remain static in the presence of a horizontal temperature gradient \citep{land}.
Any differential heating causes pressure differences in the upper atmosphere to arise due to the larger exponential scale height
(pressure gradient by altitude) of a warmer versus a colder atmospheric column.
As illustrated by the Bjerknes circulation theorem \citep{thorpe03}, this implies circulation:
upper-level air divergence from the warmer air column and low-level convergence towards it.
However, to estimate the strength of this circulation requires a shift from qualitative to quantitative considerations.

The magnitude of the surface pressure gradient can be found if one knows the isobaric height --
– the altitude where pressure does not vary over space.  Where temperature is high (and air density is low) there is less air below the isobaric height
than where temperature is low (and air density is high) -- this follows from the hypsometric equation, which captures the hydrostatic equation
and the ideal gas law.  Accordingly, the weight of the air column is lower in warmer than in colder areas. The resulting surface pressure and
temperature gradients can be shown to be proportional to each other, with the proportionality
coefficient depending on the isobaric height.

It appears that if we could determine isobaric height from some independent considerations we could use air temperature to predict surface pressures.
To address this challenge \citet{bayr13} proposed a simple physical model of temperature-driven air
redistribution which they claimed to satisfactorily quantify the relationship between tropical sea level pressure $p_s$
and the mean tropospheric temperature $T_a$ under the assumption that the isobaric height $z_e$
is the height of the troposphere, $z_e = 16.5$~km.
However, while not cited by \citet{bayr13}, in an earlier influential study \citet{lindzen87}
suggested that a similar relationship between $p_s$ and surface temperature $T_s$ can be quantitatively explained
assuming that the tropical isobaric height is around 3~km. This discrepancy requires a discussion,
because with $z_e =3$~km the model of \citet{bayr13} no longer agrees with observations.

Another point is that to obtain a satisfactory agreement with the data in their model \citet{lindzen87}
had to use an additional parameter besides the isobaric height. This additional parameter describes how fast surface temperature differences diminish
with altitude thereby defining a certain approximately isothermal height where no information about the surface temperature contrasts is preserved.
This isothermal height describes horizontal variaton in the vertical lapse rate of air temperature.
Mean tropospheric temperature investigated by \citet{bayr13} should be clearly affected by such variation.
\citet{bayr13} neglected any spatial variability in lapse rate in their model.

Here we re-examine the derivation of \citet{bayr13} to identify and resolve several inconsistencies in their model (Section~\ref{bayr}).
We derive a general relationship linking the ratios of horizontal differences
of surface pressure and air temperature (surface and mean tropospheric) to an isobaric height. We show that,
in agreement with the model of \citet{lindzen87} and in contrast with the model of \citet{bayr13}
who consider only isobaric height, these ratios are a function
of two heights, isobaric and isothermal (Section~\ref{ih}).

Using data provided by the National Centers for Environmental Prediction/National Center for Atmospheric Research (NCEP/NCAR) Reanalysis \citep{kalnay96}
and the Remote Sensing Systems \citep{mears09} we then assess the isobaric and isothermal heights in the tropics
(Section~\ref{met}).
We demonstrate in theory that the relationship between
sea level pressure and $T_a$ is significantly more sensitive to any changes in isothermal height than is the relationship between $p_s$ and $T_s$.
Accordingly, the ratio $\Delta p_s/\Delta T_a$ is not constant in the tropics and increases by about a factor of three from the higher latitudes
towards the equator. Meanwhile the $\Delta p_s/\Delta T_s$ is more spatially stable (Section~\ref{spa}).
Our analysis of the data further reveals that neither of the two assumptions made by \citet{lindzen87} concerning the isobaric and
isothermal heights appears plausible. The isobaric height is highly variable with a different distribution
for land and ocean. The isothermal height is also spatially variable.
We show that this variability does not allow one to estimate the relationship between sea level pressure and temperature
from the values of isobaric and isothermal heights with any certainty (Section~\ref{ihdata}).

We then discuss temporal versus spatial variability of sea level pressure and air temperature
(Section~\ref{tem}). While this distinction was not clearly drawn
by \citet{bayr13}, we show that their data reveal an interesting pattern.
The observed long-term temporal changes of sea level pressure and mean tropospheric temperature
are characterized by the same ratio as their respective spatial changes. In both cases land displays a smaller by absolute magnitude ratio
than the ocean. This pattern matches the observations but is not reproduced in the multimodel ensemble of the Intergovernmental
Panel on Climate Change (IPCC).

We conclude with a discussion of possible directions for future research
as to how consideration of the relationships between sea level pressure and air temperature could inform our understanding of
the principles governing low-level atmospheric circulation and moisture convergence (Section~\ref{dis}).

\section{The model of \citet{bayr13}}
\label{bayr}

\citet{bayr13}  begin their derivation with an equation they refer to as "the hydrostatic equation"
\beq\label{b1}
dp = -\rho g d\eta
\eeq
with pressure $p$, density $\rho$, gravity constant $g$, and $\eta$
described as "air column height"\footnote{In the derivation of \citet{bayr13} $\eta$ in (\ref{b1}) is denoted as $h$.}.
According to \citet{bayr13}, for an "isobaric thermal expansion of the air column"
it follows from the ideal gas law that
\beq\label{b2}
d\eta = \frac{\eta}{T}dT,
\eeq
where $T$ is temperature. They conclude that using
Eqs.~(\ref{b1}) and (\ref{b2}) one obtains how sea level pressure depends on temperature
\beq\label{b3}
\frac{dp}{dT} = \frac{1}{2}\rho g \frac{\eta}{T}.
\eeq

\begin{figure*}[h]
\centerline{
\includegraphics[width=0.90\textwidth,angle=0,clip]{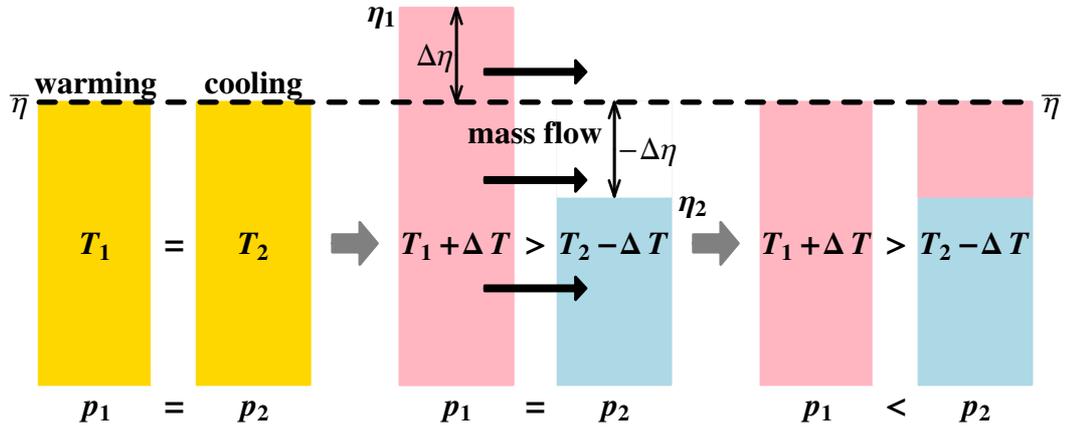}
}
\caption{\label{figcol}
Schematic of the physical model of \citet{bayr13} with a constant
mean column height $\overline{\eta}$ (re-drawn from Fig.~4 of \citet{bayr13}).
To balance two columns of heights $\eta_1$ and $\eta_2$,
one should either move to the second column from the first {\it half} of the total height difference $\eta_1 -\eta_2$ or
the {\it entire} difference $\Delta \eta \equiv \eta_1 - \overline{\eta}$ between $\eta_1$ and the mean height $\overline{\eta} \equiv (\eta_1 + \eta_2)/2$.
}
\end{figure*}

We first note that both the $1/2$ multiplier and the lack of the minus sign in (\ref{b3})
are not consistent with (\ref{b1}) and (\ref{b2}).
\citet{bayr13} explain the appearance of $1/2$ using a graphical scheme that we have re-drawn in Fig.~\ref{figcol}.
They explain that "to balance the heights of the two columns at the end,
half of the height difference is moved from the warmer to the colder air volume". As we can see from Fig.~\ref{figcol}, this statement refers
to the difference in heights $\eta$ between two local columns. However,
to test their model against the data, \citet{bayr13} define
$dp$ and $dT$ in (\ref{b3}) to represent the relative changes $d(p_s - \overline{p_s})$ and $d(T_a - \overline{T_a})$,
where $p_s$ and $T_a$ are the local values of sea level pressure and mean tropospheric temperature, respectively,
and the overbar their mean values in the tropics.
If the mean values $\overline{\eta}$, $\overline{p_s}$ and $\overline{T_a}$ change negligibly in time compared to
their local values (i.e. $d\overline X \approx 0$ with $X = p_s, T_a, \eta$), replacement of variables $X \to X - \overline X$ does not affect
Eqs.~(\ref{b1}) and (\ref{b2}). But it does impact the balancing procedure in Fig.~\ref{figcol}.
Indeed, to balance height $\eta$ between the two columns
one has to move not one half but the {\it entire} difference $d(\eta - \overline{\eta})= (1/2) (\eta_1 - \eta_2)$ from the warmer to the colder column.
Therefore, if by $dp/dT$ in (\ref{b3}) one understands, as do \citet{bayr13}, relative changes of the respective variables, one has no grounds
to introduce the $1/2$ multiplier into Eq.~(\ref{b3}) (see Eq.~(\ref{Ze1}) in the next section).

The sign discrepancy between (\ref{b1}) and (\ref{b3}) appears as a simple error, but
in fact it manifests the misapplication of Eq.~(\ref{b1}). This equation is not a hydrostatic equation and contradicts the latter.
Let us illustrate this point. For atmospheric air conforming to the ideal gas law
\beq\label{ig}
p = N R T,\,\,\,R = 8.3~{\rm J~mol^{-1}~K}^{-1},
\eeq
where $N$ is molar density, the hydrostatic equilibrium equation is
\beq\label{he}
\frac{\pt p(z)}{\pt z} = -\rho(z) g = -\frac{p}{h},\,\,\,h\equiv \frac{RT(z)}{Mg},
\eeq
where $M$ is molar mass and $z$ is height above the sea level, $p$, $\rho$ and $T$ are local
values of pressure, density and temperature at height $z$. The hydrostatic equilibrium equation (\ref{he})
says nothing about temporal changes of either pressure or temperature. It only describes the
distribution of air pressure with height.

In Eqs.~(\ref{b1})-(\ref{b3}) \citet{bayr13}
interpreted pressure $p$ as sea level pressure $p = p_s$, density $\rho$
as the mean air density in the troposphere $\rho = \rho_a = 0.562$~kg~m$^{-3}$
and air column height $\eta$ as the height $H = 16.5$~km of the tropical troposphere corresponding to height $\eta_{100}$ of pressure level $p_{100}=100$~hPa.
They also interpreted differentials in (\ref{b1})-(\ref{b3}) as describing temporal changes of the corresponding variables.
From (\ref{he}) we find that sea level pressure $p_s$ is related to $\rho_a$ and $\eta_{100}$ as follows:
\beq\label{he100}
p_s \equiv \int_0^\infty \rho g dz = \rho_a g \eta_{100} + p_{100}, \quad \rho_a \equiv \frac{1}{\eta_{100}}\int_0^{\eta_{100}} \rho(z) dz.
\eeq
Taking differential of (\ref{he100}) we obtain
\beq\label{dhe}
dp_s = \rho_a g d\eta_{100} + g \eta_{100} d\rho_a.
\eeq
When we compare (\ref{dhe}) and (\ref{b1}) with $p = p_s$, $\rho = \rho_a$ and $\eta = \eta_{100}$ it is apparent that in (\ref{b1})
the minus sign was incorrectly added to the first term in (\ref{dhe}). The second term in (\ref{dhe}), compressibility of the atmospheric air $d\rho_a \ne 0$, was
dropped altogether.

For an incompressible fluid with $\rho = \rho_a = constant$ Eq.~(\ref{dhe}) has
the familiar meaning relating column height to surface pressure: the larger the height of the fluid column,
the higher the surface pressure. Meanwhile the minus sign in Eq.~(\ref{b1}) presumes exactly the opposite:
the larger the column height $\eta$, the smaller the surface pressure.
On the other hand, if \citet{bayr13}, having ignored air compressibility for an unknown reason, used Eq.~(\ref{b1}) with the plus sign,
the sign discrepancy between (\ref{b3}) and (\ref{b1}) would have disappeared. However, in this case Eq.~(\ref{b3}) would have yielded a positive relationship of surface pressure on temperature (higher
pressure at larger temperature) which contradicts the observations: in the real atmosphere lower sea level pressure is associated with higher temperature.

We conclude that as it is based on an incorrect equation (\ref{b1}) the model
of \citet{bayr13} lacks any credible quantitative explanatory power and
yields values similar to observations only by chance.

\section{Dependence of sea level pressure on temperature}
\label{ih}

We will here derive a general relationship linking surface pressure and
temperature to the vertical structure of the atmosphere.
The model of \citet{bayr13} did not consider how temperature might vary with height.
We will allow air temperature to vary with height with a lapse rate $\Gamma \equiv -\pt T/\pt z$, which is
independent of height but can vary in the horizontal direction.

We introduce the following dimensionless variables to replace height $z$ and lapse rate $\Gamma$:
\beq\label{not}
Z \equiv \frac{z}{h_s},\,\,\,c \equiv \frac{\Gamma}{\Gamma_g},\,\,\,h_s \equiv \frac{RT_s}{Mg} \equiv \frac{T_s}{\Gamma_g},\,\,\,
\Gamma_g \equiv \frac{Mg}{R}=34\,{\rm K\,km}^{-1},\,\,\,M = 29~{\rm g~mol}^{-1}.
\eeq
Here $\Gamma_g$ is the so-called autoconvective lapse rate.
For air temperature we have
\beq\label{T}
T(Z) =T_s(1 - c Z),\,\,\,Z < c^{-1},\,\,\,T_s \equiv T(0).
\eeq
In these variables the hydrostatic equilibrium equation (\ref{he}) assumes the form
\beq\label{heG}
-\frac{\pt p}{\pt Z} = \rho g h_s = \frac{p}{1-c Z}.
\eeq
Solving (\ref{heG}) for $p \ge 0$ we have
\beq\label{heG1}
\ln\frac{p}{p_s} = -\int_0^Z \frac{dZ'}{1-c Z'}= \frac{1}{c}\ln(1-c Z) \approx
- Z - \frac{1}{2}c Z^2.
\eeq
The approximate equality in (\ref{heG1}) holds for $c Z \ll 1$, which for $\Gamma \approx 6$~K~km$^{-1}$
corresponds to $z \ll h_s (\Gamma_g/\Gamma) \approx 50$~km. This is always the case in the troposphere.

Pressure $p(z)$ and temperature $T(z)$ at a given height $z$ are functions of $p_s$, $T_s$ and $\Gamma$.
Considering linear deviations from the mean values of $\overline{p_s}$, $\overline{T_s}$,
and $\overline{\Gamma}$ and taking the total differential
of the approximate relationship for $p$ (\ref{heG1}) over these three variables we obtain:
\beq\label{dp}
dp = p_s (da + Zdb - \frac{1}{2}Z^2dc) e^{-Z},
\eeq
where $da$, $db$ and $dc$ stand for the dimensionless differentials of $p_s$, $T_s$ and $\Gamma$:
\beq\label{diff}
da \equiv \frac{dp_s}{p_s} \approx \frac{dp_s}{\overline{p_s}},\,\,\,
db \equiv \frac{dT_s}{T_s}\approx \frac{dT_s}{\overline{T_s}},\,\,\,
dc \equiv \frac{d\Gamma}{\Gamma_g},
\eeq
where $\overline{p_s} = 1013$~hPa and $\overline{T_s} = 298$~K are the annual mean
sea level pressure and surface air temperature in the tropics.
The inaccuracy of the approximate relationships in (\ref{diff}) is determined by the
relative changes of sea level pressure and surface temperature across the tropics. For the zonally averaged $p_s$
and $T_s$ this inaccuracy does not exceed $4\%$.

Isothermal height $z_i \equiv Z_i h_s$ is found by taking
total differential of $T$ (\ref{T}) over $T_s$ and $\Gamma$ and putting $dT = dT_s - T_s Z_i dc = 0$. This gives
\beq\label{Zi}
Z_i = \frac{db}{dc} = \frac{1}{h_s}\frac{dT_s}{d\Gamma},\,\,\,z_i \equiv Z_i h_s = \frac{dT_s}{d\Gamma}.
\eeq
We can see from (\ref{Zi}) that an isothermal height exists if only $db/dc > 0$, i.e.
if areas with a warmer surface have a higher lapse rate. \citet{bayr13} note that this pattern should
be related to moisture availability. They note that, below an isothermal surface, drier areas should have a steeper lapse rate
close to dry adiabat and thus get warmer than moist areas where the lapse rate is lower because of latent heat release.
In the tropical atmosphere, moist areas, most notably the equatorial regions,
have on average a steeper mean tropospheric lapse rate than do the drier regions at higher tropical latitudes (Fig.~\ref{figlap}).
In the lower atmosphere this has to do with the trade wind inversion, also mentioned by \citet{bayr13}, which is mostly pronounced
in the drier (colder) regions where the lapse rate in the low atmosphere is very small. In the upper troposphere (around the isothermal height)
a higher lapse rate in the moister regions is due to the fact that in such regions the air ascends rapidly and thus
has a lapse rate more close to adiabatic than in the slowly descending air, where a more significant part of the thermal energy can be radiated to space.
It is only in the middle atmosphere that, because of latent heat release, the lapse rate over the moist
equatorial areas is smaller than it is at higher tropical latitudes. Generally, both in the lower
atmosphere and on average in the troposphere $db/dc >0$ is fulfilled.

Isobaric height $z_e \equiv Z_e h_s$ is defined from (\ref{dp}) as the height where $dp = 0$.
It is determined from the following quadratic equation:
\beq\label{Ze}
da + Z_e db - \frac{1}{2} Z_e^2 dc = 0, \,\,\,Z_e = Z_i\left(1 \pm \sqrt{1+\frac{2}{Z_i} \frac{da}{db}}\right).
\eeq
There can be two isobaric heights (Fig.~\ref{figdiff}).
Note that the isobaric height $Z_e$ (\ref{Ze}) does not
depend on lapse rate $c$ but only on its differential $dc$ via $Z_i$. This is a consequence
of the smallness of $c Z \ll 1$ in the troposphere.

From (\ref{Ze}) and (\ref{Zi}) we obtain the following relationship for the
ratio of the differentials of surface pressure and temperature (\ref{diff}):
\beq\label{a/b}
\frac{da}{db} = -Z_e \left(1-\frac{1}{2} \frac{Z_e}{Z_i}\right).
\eeq

When $db = 0$, i.e., when the surface temperature does not vary, but only lapse rate does,
we have from (\ref{Ze})
\beq\label{a/c}
\frac{da}{dc} = \frac{Z_e^2}{2}.
\eeq
The surface pressure change is proportional to the change in lapse rate, i.e. the pressure is lower
where the lapse rate is smaller, with the proportionality coefficient equal to half the squared isobaric height.

To find the relationship between the isobaric height and mean tropospheric temperature $T_a$
below the 100~hPa level we need to find the relationship between $T_a$ and $T_s$. This relationship takes
the form (see Appendix):

\beq\label{dn}
\frac{db}{dn} = \frac{1}{1- 0.66/Z_i},\,\,\,dn \equiv \frac{dT_a}{T_a}.
\eeq

Finally from (\ref{a/b}) and (\ref{dn}) we obtain:
\beq\label{a/n}
\frac{da}{dn} = -Z_e \left(1-\frac{1}{2} \frac{Z_e}{Z_i}\right)\frac{1}{1-0.66/Z_i}.
\eeq
Relationships (\ref{a/b}) and (\ref{a/n}) are shown in Fig.~\ref{figteor}.

When, as in the model of \citet{bayr13}, lapse rate is assumed to be constant with $dc = 0$,
we have $Z_i = \infty$
and (\ref{a/n}), using notations (\ref{diff}) and $dn \equiv dT_a/T_a$, becomes
\beq\label{Ze1}
\frac{da}{db} = \frac{da}{dn} = -Z_e,\,\,\,\,\,\frac{dp_s}{dT_a}=-\frac{z_e}{h_s}\frac{p_s}{T_a} = -\rho_s g\frac{z_e}{T_a}.
\eeq
Comparing (\ref{Ze1}) to (\ref{b3}) of \citet{bayr13} we notice the absence of coefficient $1/2$ in (\ref{Ze1})
and the presence of surface air density $\rho_s$ in (\ref{Ze1}) instead of mean tropospheric air density $\rho = \rho_a$
in (\ref{b3}). If the lapse rate did not vary in the horizontal plane,
Eq.~(\ref{Ze1}) would be the correct equation relating ratio of pressure and temperature differences
to an isobaric height. But as we will show below in the real atmosphere the lapse rate variation cannot be neglected.

Considering $dp = \Delta p(z)$ in (\ref{dp}) as a small pressure difference at a given height between two air columns,
we note that this difference has a maximum above the isobaric height $Z_e$ (\ref{Ze}) at a certain height $Z_0$. This height is determined
by taking the derivative of (\ref{dp}) over $Z$ and equating it to zero, see (\ref{dp}), (\ref{Ze}) and (\ref{Zi}):
\beq\label{Z0}
\frac{\pt \Delta p}{\pt Z} = 0,\,\,\,da+Z_0db -\frac{1}{2}Z_0^2dc - db + Z_0dc =0,\,\,\,
Z_0 = 1+ Z_i \pm \sqrt{(Z_e-Z_i)^2+1}.
\eeq
At this height the pressure difference is equal to
\beq\label{dp0}
\Delta p_0 \equiv \Delta p(Z_0) = p_s e^{-Z_0} \left(da + Z_0db -\frac{1}{2}Z_0^2dc\right)=p_s e^{-Z_0} (db - Z_0 dc).
\eeq
Note that by definition when $\Delta p_0 = 0$ we have $Z_e = Z_0 = Z_i$.
As is clear from Fig.~\ref{figdiff},
where the vertical profiles of $\Delta p(z)$ (\ref{dp}) are shown for different
values of $da$, $db$ and $dc$, $\Delta p_0$ is the maximum pressure difference
between the air columns above the lower isobaric height.

As follows from Eqs.~(\ref{Z0}), (\ref{dp0}) and (\ref{a/b}), height $Z_0$ as well as the ratio between the pressure surplus aloft and
the pressure shortage  at the surface $\Delta p_0/\Delta p_s$ are functions of two parameters, the isobaric and isothermal heights $Z_i$ and $Z_e$.
When $Z_i$ and $Z_e$ are constant, the ratio between the pressure surplus aloft and the pressure shortage at the surface in
the warmer column is constant as well: the larger the pressure surplus aloft, the larger the surface pressure shortage,
with a direct proportionality between the two. This is consistent with the
conventional thinking about differential heating, that the upper pressure surplus causes air to diverge from
the warmer column, the total amount of gas will diminish and there appears a shortage of pressure
at the surface $\Delta p_s < 0$ \citep[e.g.,][Fig.~2]{pielke81}.

When the vertical lapse rate is constant, $dc =0$, from (\ref{Z0}) we have $Z_0 = 1 -da/db$.
In this case, as is clear from (\ref{dp0}), for small values of $da/db \ll 1$ the magnitude of $\Delta p_0$ does not depend on $da$,
but is directly proportional to $db$, i.e. to $\Delta T_s$ (\ref{diff}) (Fig.~\ref{figdiff}a).
This means that under these particular conditions a surface temperature gradient directly determines the pressure gradient {\it in the upper atmosphere.}
In this sense there is no difference between surface temperature gradient and a gradient of lapse rate related to latent heat release
-- both can only determine a pressure surplus aloft, cf.~Fig.~\ref{figdiff}a,b.
We emphasize that while under certain assumptions the magnitude of the tropospheric pressure gradient
can be approximately specified from considerations of the hydrostatic balance and air temperature gradients alone,
the magnitude of the surface pressure gradient cannot.

Generally, ratios $da/db$ and $db/dc$ in (\ref{Ze}) and (\ref{Zi})
can be understood as the ratios of the gradients of the corresponding variables, e.g.
$da/db = (\pt p_s/\pt y)/(\pt T_s/\pt y)(T_s/p_s)$, where $(\pt p_s/\pt y)/(\pt T_s/\pt y)$
is the ratio of sea level pressure and surface temperature gradients in a given $y$ direction (e.g. along the meridian). In this case for any $y$ the
value of $z_e$ (or $z_i$) has the meaning of a height where  $\pt p/\pt y = 0$ (or $\pt T/\pt y = 0$), i.e. where pressure
(or air temperature) does not vary over $y$.
These ratios can also be understood as the ratios of small finite differences between pressure
or temperature in a given grid point and a certain reference value of pressure or temperature,
$dp_s/dT_s = \Delta p_s/\Delta T_s$. This approach was taken by \citet{lindzen87} and \citet{bayr13}.
We can now estimate all parameters in (\ref{a/n}) from empirical data to compare them
with model assumptions.

\section{Data}
\label{met}

We used NCAR-NCEP reanalysis data on sea level pressure and surface air temperature, as well as on geopotential height and air temperature at
13 pressure levels provided by the NOAA/OAR/ESRL PSD, Boulder, Colorado, USA, from their Web site at \\
\url{http://www.esrl.noaa.gov/psd/} \citep{kalnay96}.
As an estimate of the mean tropospheric temperature we took TTT (Temperature Total Troposphere) MSU/AMSU satellite data
provided by the Remote Sensing Systems from their Web site at \\
\url{http://www.remss.com/measurements/upper-air-temperature} \citep{mears09}.
Monthly values of all variables were averaged over the time period from 1978 (the starting year for the TTT data) to 2013
to obtain 12 mean monthly values and one annual mean for each variable for each grid point on a regular $2.5\degree \times 2.5\degree$
global grid.\footnote{TTT data array contains 144 (360/2.5) longitude and 72 (180/2.5) latitude values each pertaining to
the center of the corresponding grid point. NCAR-NCEP data arrays contain 144 longitude and 73 latitude values each pertaining to the border
of the corresponding grid point. E.g., the northernmost latitude in the NCAR-NCEP data is 90$\degree$N, while for the TTT data it is
$90 - 2.5/2 = 88.75\degree$N. This discrepancy was formally resolved by adding an empty line to the end of the TTT data
such that the number of lines match and matching $i, j$ grid points in the two arrays. In the result, every TTT value
refers to a point in space that is 1.25 degree to the South and to the East from the coordinate of the corresponding
NCAR-NCEP value. This relatively small discrepancy did not appear to have any impact on any of the resulting quantitative conclusions (i.e. if instead
one moves TTT points to the North, the results are unchanged).}
The following pressure levels covering the tropical troposphere were considered:
1000, 925, 850, 700, 600, 500, 400, 300, 250, 200, 150, 100 and 70 hPa.
Meridional gradients $\pt X/\pt y$ of variable $X$ ($X = p_s,\, T_s$) at latitude $y$ were determined as
the difference in $X$ values at two neighboring latitudes and dividing by 2.5$\degree$:
$\pt X(y)/ \pt y \equiv [X(y+1.25\degree)-X(y-1.25\degree)]/2.5\degree$.
Local pressure differences corresponding to pressure level $p_j$ were calculated from the geopotential height differences
$\Delta p_j = (z_j - \overline{z_j}) p_j/h_j$, where $z_j$ is the local geopotential height of pressure level $p_j$,
$\overline{z_j}$ is its mean value in the considered spatial domain, $h_j = RT_j/(Mg)$ is the local
exponential pressure scale height (\ref{he}) and $T_j$ is local air temperature at this level.

\section{Spatial patterns}
\label{spa}

Our regression of the annual mean values of $\Delta p_s \equiv p_s - \overline{p_s}$ on $\Delta T_a \equiv T_a - \overline{T_a}$
(the overbars denote spatial averaging) for the tropical area between 22.5$^{\rm o}$S and 22.5$^{\rm o}$N produced a slope of
$r=-2.3$~hPa~K$^{-1}$ with $R^2 = 0.75$  (Fig.~\ref{figplot}a). This is practically identical to the results of Fig.~3 of
\citet{bayr13}, where
$\Delta p_s$ and $\Delta T_a$ values for the four seasons are plotted together. The resulting regression slope $r=-2.4$~hPa~K$^{-1}$
with $R^2 = 0.76$ was interpreted by \citet{bayr13} as describing seasonal changes of sea level pressure and tropospheric temperature.
However, as our result shows, even if seasonal changes of $\Delta p_s/\Delta T_a$ were completely absent,
the corresponding regression for the four seasons combined would nevertheless produce a non-zero slope reflecting the time-invariable
spatial association between higher temperature and lower pressure in the tropics. The agreement between our
relationship for the annually averaged $\Delta p_s$ and $\Delta T_a$ ratio and the one shown in Fig.~3 of
\citet{bayr13} indicates that either the spatial variation dominates the seasonal changes or that the seasonal changes are, on average, characterized
by a similar $\Delta p_s/\Delta T_a$ ratio as the spatial changes. \citet{bayr13}
did not discuss whether their Fig.~3 actually characterizes spatial or temporal variation but
interpreted the results of their Fig.~3 as a test of validity of their model which they later used to assess long-term temporal changes in $p_s$ and $T_a$.

Next we note that since a linear regression minimizes the departure of the empirical points from the theoretical line,
the regression slope can be disproportionately influenced by the values that depart most from the mean. This
depends on the shape of the frequency distribution of data points around the mean. For
data points in Fig.~\ref{figplot}a with their $\Delta p_s$ departing from the zero mean by more than two thirds of standard deviation
the regression slope is $-2.5$~hPa~K$^{-1}$ with $R^2 = 0.85$, which is very close to the overall regression (cf. Fig.~\ref{figplot}a,b).
Meanwhile the regression slope for the remaining grid points with smaller $|\Delta p_s|$
is $-1.4$~hPa~K$^{-1}$ with $R^2 = 0.30$ (Fig.~\ref{figplot}c). This subset constitutes about half of all the points but harbors
two thirds of all land values (Fig.~\ref{figplot}d). This subset apparently makes a negligible contribution to the pantropical regression which,
in consequence, appears uninformative with regard to a large portion of the data, including most land.

We further found that the regression slope $r$ depends strongly on the averaging domain: it increases by absolute magnitude towards the equator
(Fig.~\ref{figgra}a). As $r$ decreases with diminishing
tropical area, so does the squared correlation coefficient (Fig.~\ref{figgra}b),
although for the oceanic grid points it remains relatively high even near the equator. E.g., for the area between 30$^{\rm o}$S and 30$^{\rm o}$N
for the ocean we have
$r =-1.5$~hPa~K$^{-1}$ with $R^2 = 0.75$ while for 5$^{\rm o}$S and 5$^{\rm o}$N  we have $r =-4.2$~hPa~K$^{-1}$ with $R^2 = 0.64$.
These patterns testify that the spatial relationship
between $\Delta p_s$ and $\Delta T_a$ is not universal in the tropics. In contrast, while similar regressions of $\Delta p_s$ on
surface temperature $\Delta T_s$ are characterized by lower $R^2$, the regression slope, around $-1$~hPa~K$^{-1}$, does not appear to depend
significantly on the averaging domain (Fig.~\ref{figgra}d,e).

To explore whether the relationship between $\Delta p_s$ and $\Delta T_a$ is dominated by zonal or meridional differences, we
performed a regression of zonally averaged $\Delta p_s$ on zonally averaged $\Delta T_a$ for the area between 22.5$^{\rm o}$S and 22.5$^{\rm o}$N
for different months (Fig.~\ref{figgra}c). Zonally averaged values account for a major part of the dependence between $\Delta p_s$ and $\Delta T_a$:
regression of annual mean zonally averaged values yields $r=-2.0$~hPa~K$^{-1}$. The same is true for the surface temperature (Fig.~\ref{figgra}f).
Since land/sea contrasts in the tropics are predominantly zonal, this means, again, that either the contribution of land/sea pressure/temperature
contrasts
to the pantropical regressions of $\Delta p_s$ on $\Delta T_a$ and $\Delta T_s$ is relatively unimportant or that these contrasts
are characterized by approximately the same ratio as the zonally averaged values.

\section{Isobaric and isothermal heights}
\label{ihdata}

While the models of \citet{lindzen87} and \citet{bayr13} build upon the notion of an isobaric height,
neither study provided a systematic assessment of the observational evidence to
quantify its magnitude and variation in space and time.
Given the dominance of zonally averaged patterns an initial insight into the behavior of isobaric and isothermal heights can be gained from comparing
vertical profiles of the zonally averaged pressure and temperature gradients (Fig.~\ref{figtemp}). We can see that there
is a universal pantropical isothermal height around 12~km. At the same time closer to the equator the minimal height
where the surface temperature contrasts disappear diminishes (Fig.~\ref{figtemp}a-c).
For comparison, in their model \citet{lindzen87} adopted a constant isothermal height equal to 10~km. (They assumed that the horizontal temperature
differences at the level of $z_{LN} = 3$~km are 30\% smaller than the corresponding differences
at the sea level: $\Delta T(z_{LN}) = 0.7 \Delta T_s$. From $T(z_{LN}) = T_s - \Gamma z_{LN}$ and (\ref{Zi}) we obtain
$z_i \equiv \Delta T_s/\Delta \Gamma = z_{LN}/0.3 = 10$~km.)

In agreement with Eq.~(\ref{Ze}), there are two minima of pressure gradients corresponding to two isobaric heights,
one closely above the troposphere and another one in the lower atmosphere. The height of the lower minimum grows towards the equator
(Fig.~\ref{figtemp}d-f). As the upper isobaric height is predominantly above the 70 hPa level, we cannot estimate it with certainty in our data.
However, there is a tendency for this height to slightly diminish towards the equator.

Generally, a pantropical isobaric (isothermal) height, if it exists, has the following properties: at this height (1) deviation of local
pressure (temperature) from the pantropical mean is zero; (2) deviation of local pressure (temperature) from the zonal
mean is zero; (3) local horizontal gradient of pressure (temperature) is zero.
If a universal isobaric height does not exist, for each grid point these properties can each define a different height.
In Fig.~\ref{figiso}a,d we plotted zonal averages of the minimal isobaric and isothermal heights calculated according to the above
definitions. Local vertical profiles of pressure and temperature differences in individual grid points illustrating how these heights were calculated
are exemplified in Fig.~\ref{figiso}b,e.

In the majority of grid points there is an isobaric height between 0 and 10~km (Fig.~\ref{figiso}c),
which corresponds to the lower isobaric height from Eq.~(\ref{Ze}).
It is of interest that the land and the ocean have different lower isobaric heights, with land values peaking below
the trade wind inversion layer (3~km) and ocean values peaking at around 6~km (Fig.~\ref{figiso}c).
As is clear from Fig.~\ref{figiso}f, a significant part of grid points has two isothermal heights: one
is the pantropical isothermal height around 12~km and the second one is around 3~km.

While our relationship (\ref{a/n}) has the limitation of not accounting for the vertical
changes in lapse rate, it does provide an insight into the observed behavior of the
$\Delta p_s/\Delta T_s$ and $\Delta p_s/\Delta T_a$ ratios.
From Fig.~\ref{figteor}a we can see that $\Delta p_s/\Delta T_s$ grows with increasing lower isobaric height $z_{e1}$
and, for a given $z_e$, declines with decreasing isothermal height $z_i$.
As there is a tendency for $z_{e1}$ to grow and for $z_i$ to diminish towards the equator (Fig.~\ref{figiso}a,d),
this compensating behavior may explain the approximate constancy
of $\Delta p_s/\Delta T_s$ (Fig.~\ref{figgra}d). Meanwhile
because of the singularity for $z_i \approx 6$~km, a decrease in $z_i$ for $z_e \approx 6$~km, in contrast, leads to a sharp
increase in $|\Delta p_s/\Delta T_a|$. This theoretical behavior is consistent with the observed growth of $|\Delta p_s/\Delta T_a|$
in the vicinity of the equator (Fig.~\ref{figgra}a).

Fig.~\ref{figteor} also illustrates the sensitivity of these ratios to the values of the lower and upper isobaric heights $z_{e1}$ and $z_{e2}$.
With $z_{e1}\ll z_i$ the ratio of $\Delta p_s/\Delta T_a$ grow approximately proportionally to $z_{e1}$.
Accordingly, land with its significantly lower $z_{e1}$ have lower $\Delta p_s/\Delta T_a$
and $\Delta p_s/\Delta T_s$ than the ocean (Fig.~\ref{figgra}a). Thus the observed variability in the lower isobaric height produces
uncertainties of the order of 100\% in the corresponding estimates of those ratios.
While the upper isobaric height appears more conservative, the sensitivity of $\Delta p_s/\Delta T_a$
to its value is much higher. For example, for $z_i = 10$~km a change in $z_{e2}$ of about 10\% from 18~km to 20~km
diminishes the magnitude of $|\Delta p_s/\Delta T_a|$ by more than an order of magnitude (Fig.~\ref{figteor}b).

\section{Temporal patterns}
\label{tem}

We now investigate the time dependence of the relationship between pressure and temperature. If in a certain area
we have an isobaric surface at height $z_e$ and an isothermal surface at height $z_i$, we have (see Eq.~\ref{a/n}):
\beq\label{dy}
\Delta p_s = r \Delta T_a,\,\,\, r \equiv -\frac{z_e}{h_s}\frac{p_s}{T_a} \left(1 - \frac{z_e}{2z_i}\right) \frac{1}{1-0.66h_s/z_i}
\eeq
Local values of $\Delta p_s \equiv p_s - \overline{p_s}$ and $\Delta T_a \equiv T_a - \overline{T_a}$ are defined
with respect to their mean values in the area where the isobaric and isothermal surfaces exist.
Taking the derivative of (\ref{dy}) over time we obtain
\beq\label{dt}
\frac{\pt \Delta p_s}{\pt t} = r \frac{\pt \Delta T_a}{\pt t}\,\,\,{\rm if}\,\,\,\frac{\pt r}{\pt t} = 0,\,\,\,
\frac{\pt \Delta X}{\pt t} \equiv \frac{\pt X}{\pt t} - \frac{\pt \overline{X}}{\pt t},\,X = p_s,\,T_a.
\eeq
This means that if $z_e$, $z_i$ and, hence, $r$ (to the accuracy of a few per cent) do not change with time, temporal changes of $\Delta p_s$
and $\Delta T_a$ are characterized by the same ratio $r$ as their spatial changes.

For each grid point we made a regression of the local monthly changes of sea level pressure, $\widetilde{\Delta} p_s(m)$, on
the local monthly changes of mean tropospheric temperature, $\widetilde{\Delta} T_a(m)$. A similar analysis was performed for
$p_s$ and surface temperature $T_s$. The results are shown in Fig.~\ref{figmap}.
Here $\widetilde{\Delta} p_s(m)\equiv p_s(m) - p_s(m_p)$ and
$\widetilde{\Delta} T_a(m)\equiv T_a(m) - T_a(m_p)$, where $m$ is the current month and $m_p$
is the previous month (e.g., January and December).
Thus $\widetilde{\Delta} p_s(m)/\widetilde{\Delta} m$ and
$\widetilde{\Delta} T_a(m)/\widetilde{\Delta} m$, where $\widetilde{\Delta} m = 1$~month, represent
the monthly mean temporal derivatives of sea level pressure and tropospheric temperature in a given grid point.
Linear regression of $\widetilde{\Delta} p_s$ on $\widetilde{\Delta} T_a$ characterizes the annually
averaged dependence between monthly changes of pressure and temperature
which correspond to Eq.~(\ref{dt}) when $\pt \overline{p_s}/\pt t = 0$ and $\pt \overline{T_a}/\pt t= 0$.
As we can see from Fig.~\ref{figmap}, the pantropical mean monthly changes of sea level pressure and air temperature
are significantly smaller compared to the majority of the corresponding local changes, so
conditions $\pt \overline{p_s}/\pt t = 0$ and $\pt \overline{T_a}/\pt t= 0$ approximately hold for the tropics as a whole.

Our analysis shows that in the equatorial land regions with high rainfall -- in the Amazon and Congo river basins, see point C in Fig.~\ref{figmap} -- the regressions were not
significant at $0.01$ probability level\footnote{On land, sea level pressure is not an empirically measured variable,
but is calculated from pressure $p_l(z_l)$, temperature $T_l(z_l)$ and the geopotential height $z_l$ of the land surface
assuming $\Gamma = 6.5$~K~km$^{-1}$ for $0 \le z \le z_l$, where $z = 0$ corresponds to the sea level.
This definition introduces a formal dependence of $p_{sl}$ (sea level pressure on land) on surface air temperature $T_l$, the strength
of which is directly proportional to $z_l$. That is, $p_{sl}$ diminishes with growing $T_l$ even if $p_l$ and, hence, the amount
of gas in the atmospheric column remains constant. Approximating the hydrostatic equation (\ref{he}) as $(p_l - p_{sl})/z_l = -p_l/h$, $h = RT_l/(Mg)$,
and taking the derivative of this equation over $T_l$  at constant $p_l$ we obtain $dp_{sl}/dT_l = (z_l/h) p_l/T_l$.
For the mean geopotential height $z_l = 0.6$~km of the tropical land, $p_l = 950$~hPa and $T_l = 295$~K we find
$dp_{sl}/dT_l = -0.2$~hPa~K$^{-1}$, i.e. about 20\% of the mean ratio established by us for the tropical land (Fig.~\ref{figmap}) is not related
to any air redistribution but is a formal consequence of the definition of $p_{sl}$.}. Where the regressions are significant, the largest (by absolute magnitude) regression slopes
tend to be concentrated in the regions of the largest meridional gradients of sea level pressure, i.e. around the 15 and 20 degrees latitudes (Fig.~\ref{figmap}).
These local dependences between $\widetilde{\Delta}p_s$ and $\widetilde{\Delta}T_s$ can be explained by the seasonal migration of the Hadley cells
where lower pressure is spatially associated with higher temperature (see Fig.~\ref{fighad}).

This explanation agrees with the observation that in those extratropical regions where
areas of low pressure are at the same time areas of low temperature (particularly
the southern Ferrel cell), the seasonal relationship between pressure and temperature changes is generally less consistent than it is in the tropics
with local relationships occasionally being reversed -- i.e., pressure and temperature rise or decline together (see point E in Fig.~\ref{figmap}).
Local changes in $p_s$ and $T_a$ appear affected by migration of the circulation cells with a spatially invariable temperature-pressure relationship
within the cells.

Plotting local absolute changes of $p_s$ and
$T_a$ instead of local changes relative to the tropical mean in Fig.~\ref{figmap} has the advantage that the resulting figure
does not depend on the averaging domain (cf. Fig.~\ref{figgra}a) and allows for comparison of tropical and extratropical patterns.
However, we additionally performed a regression of local relative monthly changes as in Eq.~(\ref{dt}) for each grid point
between 22.5\degree S to 22.5\degree N. The tropical mean values ($\pm$ standard deviation) of the obtained local slope
coefficients are as follows. For the relationship between $p_s$ and $T_a$:
$-2.4\pm 1.8$~hPa~K$^{-1}$ ($-2.5\pm 0.8$~hPa~K$^{-1}$ for the land, $-2.3\pm 1.9$~hPa~K$^{-1}$ for the ocean).
For the relationship between $p_s$ and $T_s$:
$-0.9\pm 0.6$~hPa~K$^{-1}$ ($-1.2\pm 0.6$~hPa~K$^{-1}$ for the land, $-1.1\pm 0.6$~hPa~K$^{-1}$ for the ocean).
We can see that these figures are again very close to the corresponding spatial ratios (Fig.~\ref{figplot}a, Fig.~\ref{figgra}a,d).
There is, however, less difference between the land and the ocean than in the spatial ratios.

In their analysis of the observed long-term $p_s$ and $T_a$ changes \citet{bayr13} compared relative partial pressures
and relative tropospheric temperatures in 1989-2010 (their Fig.~12). They found that, in agreement with Eq.~(\ref{dt}), these changes are related by
practically identical ratios,
$-2.0$~hPa~K$^{-1}$ for land, $-2.4$~hPa~K$^{-1}$ for ocean,
and $-2.3$~hPa~K$^{-1}$ for tropics as
a whole, as the corresponding spatial contrasts shown in Fig.~\ref{figplot}a. This pattern is not preserved in the IPCC multimodel ensemble \citep[][their Fig.~5]{bayr13}: modelled long-term changes
for the time period 1970-2099 are characterized by a lower ratio for land ($-2.5$~hPa~K$^{-1}$) than for the ocean ($-1.9$~hPa~K$^{-1}$)
with an overall mean of $-2.0$~hPa~K$^{-1}$.

\section{Discussion}
\label{dis}

We have critically examined the model of \citet{bayr13}. For an atmosphere where the lapse rate is everywhere the same,
the correct expression for the dependence between sea level pressure and tropospheric temperature is given by Eq.~(\ref{Ze1}):
it differs from Eq.~(\ref{b3}) of \citet{bayr13} by the absence of $1/2$ and the presence of
surface air density $\rho_s$ instead of mean tropospheric density $\rho_a$. Eq.~(\ref{Ze1}) is similar to Eq.~(\ref{b3})
in that it describes a direct proportionality between the isobaric height $z_e$ and the $\Delta p_s/\Delta T_a$ ratio.

In the real atmosphere the lapse rate varies considerably in the horizontal plane: the lapse rate over the warmer surfaces
is on average steeper than over the colder surfaces (Fig.~\ref{figlap}).
We have shown that in such an atmosphere there must be at least two isobaric heights $z_{e1} \le z_{e2}$
(Eq.~\ref{Ze}, Fig.~\ref{figdiff}c).
The models of \citet{lindzen87} and \citet{bayr13} each considered only one isobaric height,
while the existence of the second one and its link to the first was not discussed.
In agreement with our Eq.~(\ref{Ze}), observations show that in the tropical atmosphere the two isobaric heights
correspond to $z_{e1} \sim 1.5-6$~km and $z_{e2} \sim 17-20$~km.
Both heights are spatially variable (Fig.~\ref{figiso}a-c). Observations do not support
the existence of a pantropical constant isobaric height either around 3~km as in the model of \citet{lindzen87}
or at the top of the troposphere as in the model of \citet{bayr13}.
Apparently, in the presence of two isobaric heights the $\Delta p_s/\Delta T_a$ ratio cannot be
a linear function of $z_e$. Indeed, we have shown that there is an additional essential parameter governing the relationship
between pressure and temperature, the isothermal height $z_i$. The resulting dependence of $\Delta p_s/\Delta T_a$ on the isobaric
height is quadratic, not linear (Eq.~\ref{a/n}).

Isothermal height $z_i$ characterizes the thermal structure of the troposphere.
In the limit of very large $z_i$ the ratio $\Delta p_s/\Delta T_s$ (and $\Delta p_s/\Delta T_a$) do not depend
on $z_i$, but only on isobaric height $z_e$ (Eqs.~\ref{a/b}, \ref{Ze1}). If $z_e$ is given,
$\Delta p_s$ depends only on surface temperature contrasts. This fact apparently facilitated interpretation
of surface pressure contrasts as {\it determined} by surface temperature \citep{lindzen87,sobel06,an11}.
For example, \citet[][p.~324]{sobel06} in their discussion of the model of \citet{lindzen87} noted that surface temperature
determines temperature in the atmospheric boundary layer, which, in turn, determines
surface pressure via a hydrostatic relationship. Indeed, if the lapse rate does not vary in the horizontal plane,
$z_i = +\infty$ and there is a direct proportionality between $\Delta p_s$ and $\Delta T_s$ (and $\Delta T_a$).
But in the real atmosphere
$z_i$ is relatively large not because the release of latent heat in the troposphere does not matter. On the contrary, it is the latent heat release
that works to elevate the isothermal height by diminishing the difference in the mean tropospheric lapse rate between the warmer and colder
surface areas that could have otherwise been much larger (Fig.~\ref{figlap}). An illustration is shown in Fig.~\ref{figsah} which compares
pressure and temperature profiles in Sahara and East China in July each with the zonal mean profile. One can see that in the dry Sahara the isothermal
height is very small while in East China (where it is the monsoon period) it is on average much higher.
On the other hand, when $z_i$ is not very large but comparable to $z_e$, it has a crucial impact on the relationship between pressure and temperature (Eq.~\ref{a/n}).

We have shown that the character of the relationship between $\Delta p_s/\Delta T_a$ (as well as of $\Delta p_s/\Delta T_s$)
is extremely sensitive to the values of $z_e$ and $z_i$ in the interval of their observed values (Fig.~\ref{figteor}).
This sensitivity has not been previously explored. For example,
at $z_i = 10$~km, which is the value adopted by \citet{lindzen87}, a 15\% change in the upper $z_e$ from 17~km to 20~km leads to
a complete disappearance of the dependence of $\Delta p_s$ on both $\Delta T_s$ and $\Delta T_a$. Therefore, the assumed approximate constancy of
the upper isobaric height $z_{e2}$, which motivated the model of \citet{bayr13}, proves to be too poor a representation of
reality for this height to serve as a
determinant of the $\Delta p_s/\Delta T_a$ ratio. The sensitivity to the lower isobaric height $z_{e1}$ is lower: at constant $z_i$ the ratio $\Delta p_s/\Delta T_a$
is directly proportional to $z_{e1}$ (Fig.~\ref{figteor}b). However, data show that $z_{e1}$ experiences a proportionally higher spatial variation than
$z_{e2}$. It varies from a few hundred meters over land to over 6~km over the ocean (Fig.~\ref{figiso}c). In the result, $z_{e1}$ appears as invalid
for a theoretical prediction of $\Delta p_s/\Delta T_a$ as is $z_{e2}$.
With several different peaks, different values for land and ocean and
mean values depending on latitude (Fig.~\ref{figiso}a,d) the isobaric and isothermal heights can hardly be specified
from some independent physical considerations. Rather, they are themselves dictated by the dynamic relationships between pressure and temperature.

We have emphasized an important property of the isobaric height: it links sea level pressure contrasts to the pressure
contrasts in the troposphere (Eq.~\ref{dp0}). If the isobaric height is unknown, so are
the surface pressure contrasts. Unlike the tropospheric pressure contrast, the surface pressure contrasts
cannot be determined from consideration of temperature gradients alone. This summarizes a major problem for the
theory of atmospheric circulation. Having set a temperature gradient, one can easily find tropospheric pressure
gradients and, consequently, the geostrophic winds in the troposphere. However, what type of circulation can be generated
in the low-level atmosphere remains uncertain. To simulate low-level winds generated by differential heating, one has to specify the dynamic
interaction between the upper and lower atmosphere or, simply put, the turbulent friction. Depending on the adopted
parametrization, one and the same temperature gradient can be modelled to produce drastically different low-level
winds. E.g., for an axisymmetric general atmospheric circulation different assumptions regarding friction yield diverse results
from complete absence of any low-level circulation to negligible meridional circulation to circulations close to the observed \citep[e.g.,][]{held80,sc06}.
Through parametrization of turbulence, models of more local circulations driven by
differential heating adopt as granted the basic parameters of the larger-scale circulations into which they are
embedded \citep[e.g.,][]{smagor53,pielke81,curry87}. While optimized to provide an adequate description of the relevant processes, such models would yield
very different results if those basic parameters change.

A simple but relevant illustration to these ideas is provided by the model of \citet{lindzen87}. They investigated how low-level
winds depend on the isobaric height $z_e$ in their model, where the boundary layer height was assumed to be equal to $z_e$.
In the limit $z_e \to 0$ the surface pressure gradients disappear (Eq.~\ref{a/b}) and the low-level winds should vanish.
Contrary to this expectation \citet{lindzen87} found little dependence of the meridional winds (and moisture convergence) on $z_e$ in their model.
A smaller $z_e$ expectedly produced weaker surface pressure gradients, but
it also produced a proportionally larger damping coefficient $\epsilon \equiv C_D |V_c|/z_e$, where $C_D$ is a constant
and $V_c$ is a typical wind speed at $z_e$ taken to be equal to 8~m~s$^{-1}$.
As a result of a weaker meridional pressure gradient, the zonal wind did decrease proportionally to the surface pressure gradient. However, the meridional
wind proportional to the product of zonal wind and the damping coefficient $\epsilon$ \citep[][see their Eq.~12a]{lindzen87} did not change much.
Here the decrease in pressure gradient was offset by an increase in the damping coefficient $\epsilon$,
such that the low-level air convergence remained approximately independent of $z_e$ (and, hence, of surface pressure gradients).

This independence resulted from how friction was parameterized, in particular, from the assumed constancy of $V_c$
and, hence, from the inverse proportionality between the damping coefficient and the isobaric height.
In the real atmosphere the height of boundary layer $h_b$ is much smaller than the isobaric height, $h_b \sim 1\,{\rm km} \ll z_e$,
especially over the ocean (Fig.~\ref{figiso}c). Because of this, pressure gradients at the top of the boundary layer are determined by the surface
pressure gradients and close to them. Since at the top of the boundary layer winds are approximately geostrophic \citep{back09},
this means that the geostrophic wind speed $V_c$ at the top of the boundary layer (which is used in the determination of the damping coefficient)
is approximately proportional to the surface pressure gradient. Consequently, it must decrease with decreasing $z_e$. In the result, with decreasing
$z_e$ (decreasing surface pressure gradient), surface winds and moisture convergence should decline as well.
On the other hand, $z_e$ can be itself a function of pressure gradients.
This simple example illustrates that our answer to the question: "what happens in the lower atmosphere" is fully
determined by how turbulence is parameterized and is practically independent of the magnitude of temperature contrasts.
The opposite is true for the geostrophic wind in the upper atmosphere. Since turbulent friction is directly
related to the dissipative power of atmospheric circulation, finding constraints on this power will help resolve
the big challenge of predicting low-level circulation and moisture convergence \citep{jas13}.

We have seen that most part of the relationship between $p_s$ and $T_a$ in the tropics shown in Fig.~3 of \citet{bayr13}
can be attributed to the properties of the tropical atmosphere
around 20 degrees latitudes and in the region of the Walker circulation (Fig.~\ref{figplot}d)
as well as to its zonally averaged properties (Fig.~\ref{figgra}c). This means that to explain why the
ratios between sea level pressure and temperature contrasts have their observed magnitudes we have to explain why the Hadley and Walker
circulations are characterized by those pressure and temperature contrasts as they are.
Other processes, including seasonal variation and the observed long-term relative $p_s$ and $T_a$ changes as in Fig.~12 of \citet{bayr13},
appear to preserve the vertical structure of the atmosphere set by the main dynamic drivers of the circulation.
In other words, if we knew why there is a $\sim$10~hPa pressure contrast per $\sim$10\degree C tropospheric and surface temperature contrast in Hadley cells (Fig.~\ref{fighad})
we would be able to determine the isobaric heights and thus understand why the tropical troposphere is about 16~km and not, say, 10 or 25~km high
as well as why a local temperature increase leads to a local sea level pressure drop of a given magnitude.
Two drivers of low-level circulation have been considered
\citep{gill80,lindzen87,neelin89,sobel06,back09,an11}: surface heating and the release of latent heat. A distinct physical process
was recently described. Horizontal transport of moisture with its subsequent condensation and precipitation away from the point where it evaporated
produces pressure gradients due to the changing concentration of water vapor as the air moves from the evaporation to condensation area \citep{m13,mgn14}.
Pressure is greater where water vapor is added and lower where it is removed from the air column.
Understanding the relative contributions of these processes will guide our predictions of local pressure and circulation changes.
We believe that the analysis of pressure/temperature relationships initiated by \citet{bayr13}
can shed light on the relative strength of the contributing processes if these are studied
together with the moisture contrasts.

\section*{Acknowledgment}
We sincerely thank Dr. Bayr and Dr. Dommenget and the reviewers for constructive critical comments on our work.

{\clearpage}
\begin{appendix}
\section*{
\begin{center}
Relationship between $T_s$ and $T_a$
\end{center}}

The relationship between surface temperature $T_s$ and the mean temperature $T_a(Z)$ of the atmospheric column below $Z$
can be derived from (\ref{T}) and the hydrostatic equation (\ref{he}):
\beq\label{Ta}
T_a(Z) \equiv \frac{\int_0^{Z} T(Z) \rho dZ}{\int_0^{Z} \rho dZ} = \frac{T_s}{1+c}\frac{1-e^{-cZ}e^{-Z}e^{-cZ^2/2}}{1-e^{-{Z}}e^{-c {Z}^2/2}},
\,\,\,Z \equiv \frac{z}{h_s},\,\,\,cZ \ll 1.
\eeq
Expanding (\ref{Ta}) over $c$ and keeping the linear term we have
\beq\label{Ta2}
T_a = T_s\left[1 - c \left(1 - \frac{Z}{e^Z-1}\right)\right].
\eeq

\begin{figure*}[h]
\begin{minipage}[h]{0.49\textwidth}
\centerline{
\includegraphics[width=0.99\textwidth,angle=0,clip]{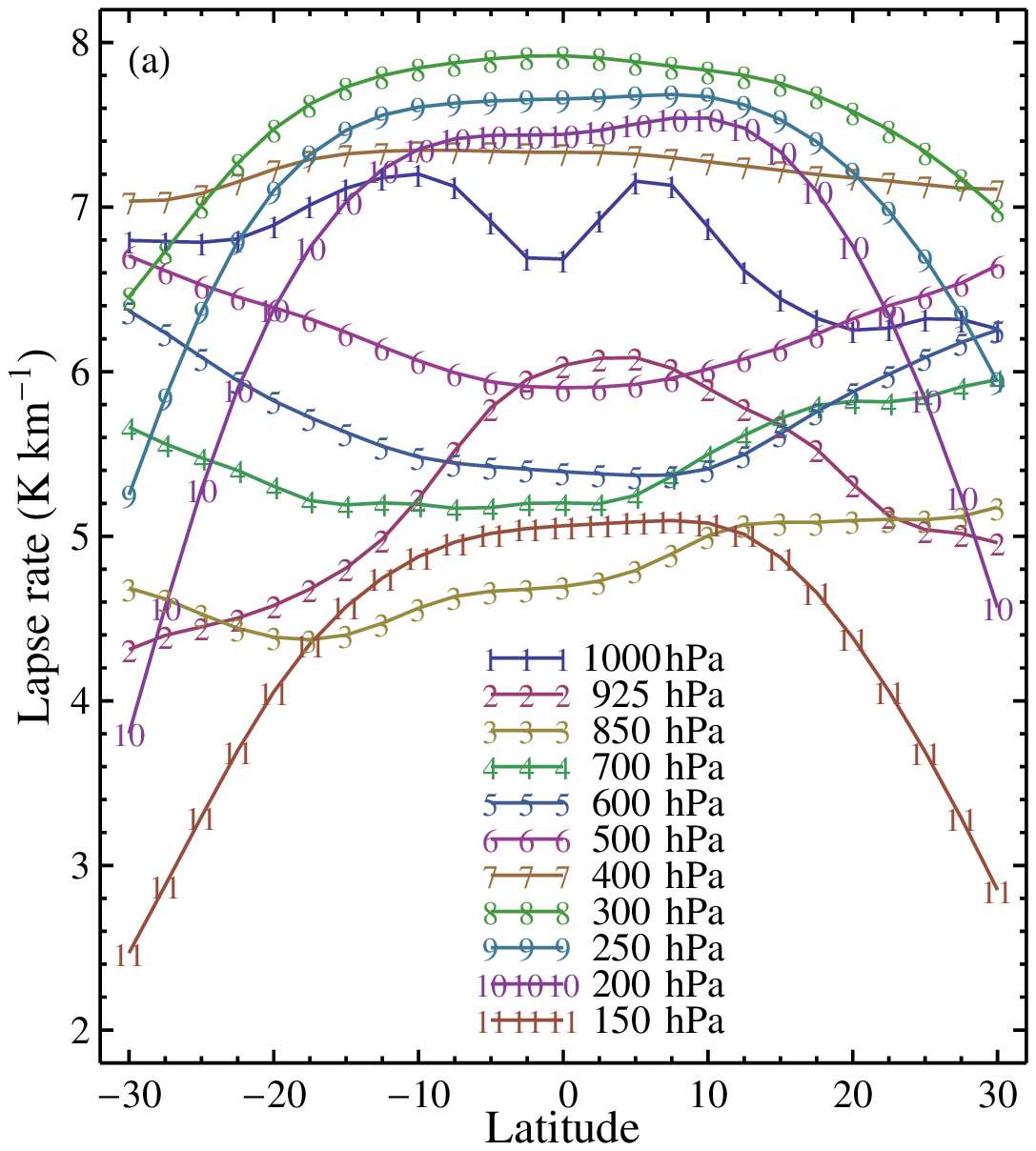}
}
\end{minipage}
\begin{minipage}[h]{0.49\textwidth}
\centerline{
\includegraphics[width=0.99\textwidth,angle=0,clip]{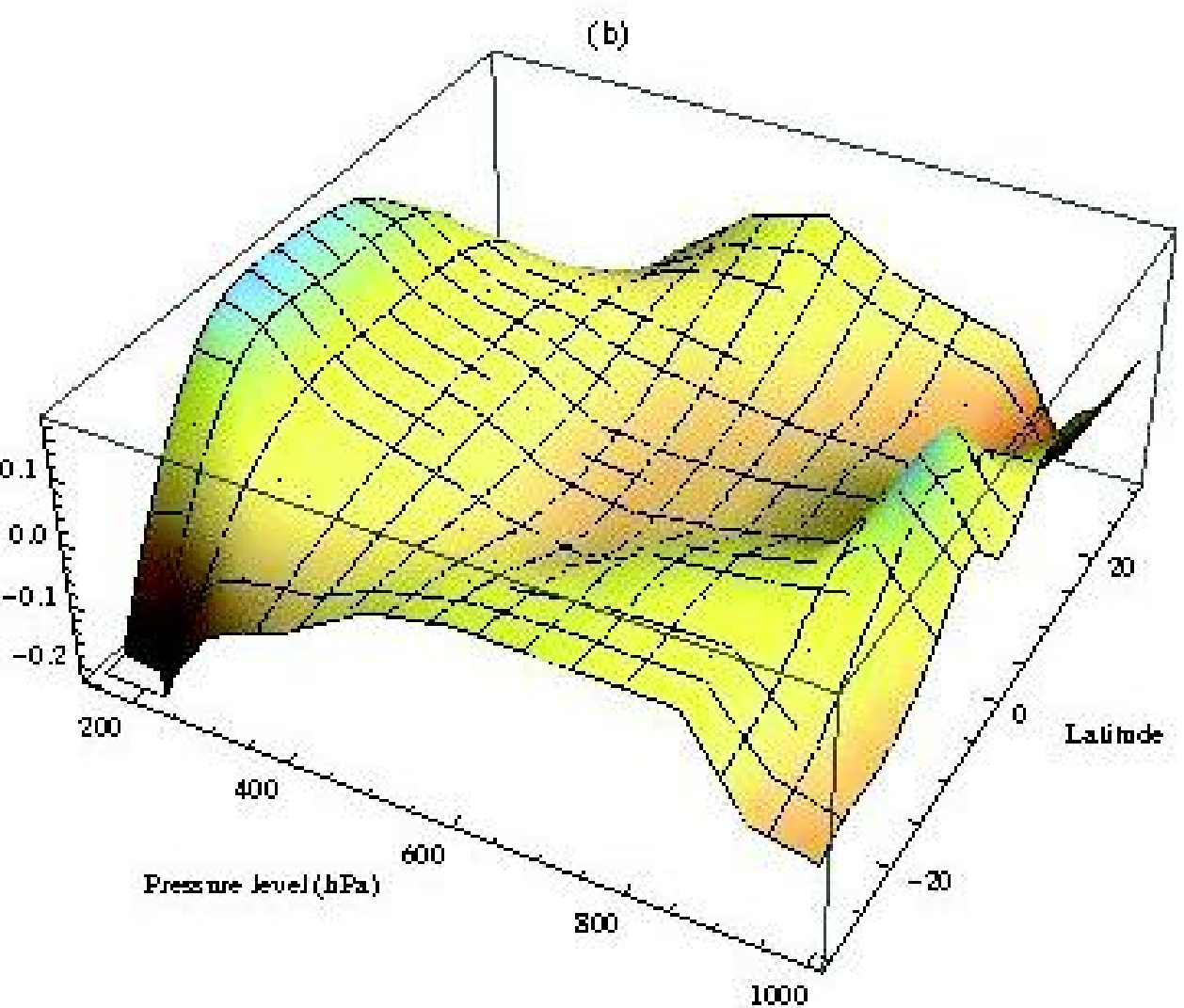}
}
\end{minipage}
\caption{\label{figlap}
Annual mean latitudinal profiles of the air temperature lapse rate on different
pressure levels. For example, curve 1 in (a) shows the mean lapse rate between 1000 hPa
and 925 hPa; curve 11 -- between 150 and 100 hPa.
The tropical mean lapse rate (the temperature difference between 1000 hPa
and 100 hPa levels divided by the difference in the geopotential heights and averaged
from 30$\degree$S to 30$\degree$N) is 6.0~K~km$^{-1}$. Panel (b) shows the relative variation -- at each pressure level the lapse rate
at a given latitude is divided by the mean lapse rate at this level (averaged between
30$\degree$S and 30$\degree$N). The equator has a higher lapse rate than the 30 degrees latitudes in the lower and upper
-- but not the middle -- troposphere.
}
\end{figure*}

\begin{figure*}[h]
\centerline{
\includegraphics[width=0.95\textwidth,angle=0,clip]{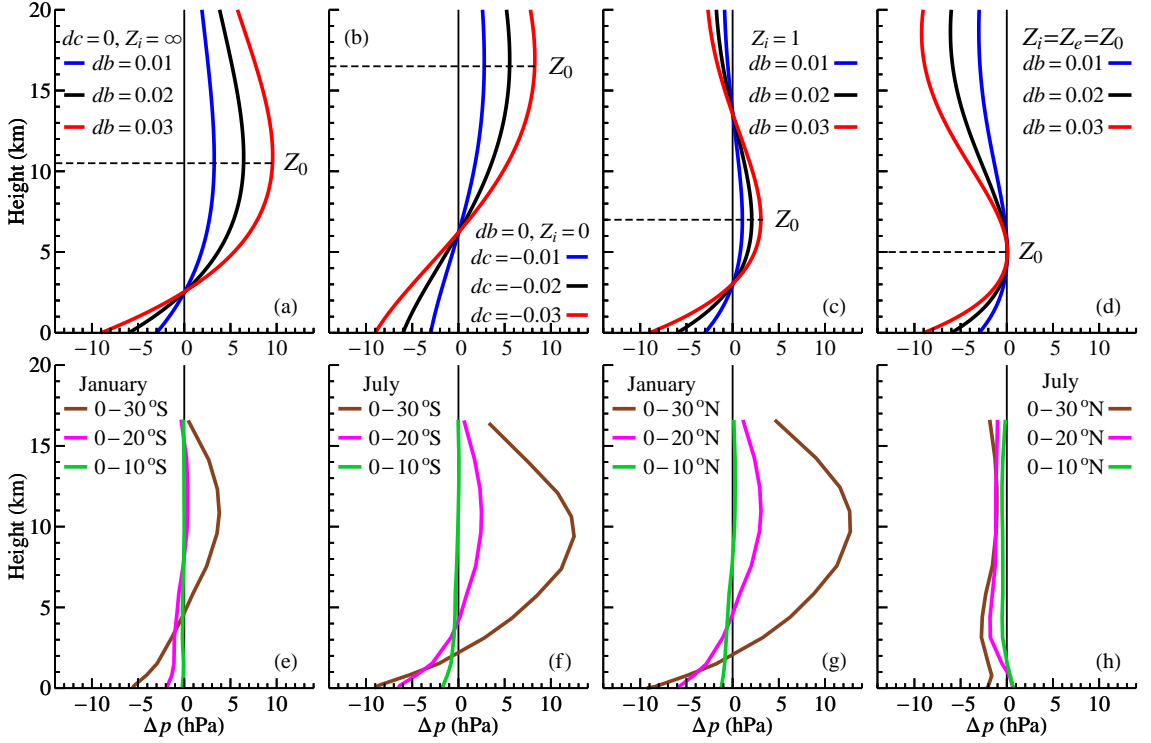}
}
\caption{\label{figdiff}
Vertical profiles of pressure differences $\Delta p(z)$ between air columns differing
in their lapse rate, surface pressure and temperature. Panels (a)-(d): theoretical
profiles (\ref{dp}) with $dp = \Delta p$, $da = \Delta p_s/p_s$,
$db = \Delta T_s/T_s$, $dc = \Delta \Gamma/\Gamma_g$ (cf. \ref{diff}), $p_s = 1000$~hPa,
$T_s = 300$~K. In panels (a)-(d) $da = -0.003$, $-0.006$, $-0.009$ for the blue, black and red curves,
respectively. In each panel $da/db = constant$ for all the three curves. Dashed line $Z_0$ (\ref{Z0}) shows the height where the positive pressure difference in the
upper atmosphere is maximum, $\Delta p(Z_0) = \Delta p_0$ (\ref{dp0}). Note two isobaric heights in
panel (c). In panel (d) condition
$Z_i = Z_0$ (the atmosphere is horizontally isothermal where the positive pressure difference aloft is maximum)
yields $Z_i = Z_0 = Z_e = -2da/db = (-2da/dc)^{1/2}$, see (\ref{Ze}), (\ref{Z0}) and (\ref{Zi}),
and $\Delta p_0 = 0$, i.e. the pressure surplus aloft disappears.
Panels (e)-(h): real vertical profiles of zonally averaged pressure differences between the air columns at the equator and
10, 20 and 30 degrees latitudes in the Southern (e,f) and Northern (g,h) hemispheres
in January (e,g) and July (f,h).
E.g., the brown line in (e) shows the difference between the air column at the equator and at 30$\degree$S in January.
Note that while the theoretical curves (a-d) in each panel are chosen such that they have one and the same isobaric height $Z_e$ (i.e.,
they cross the line $\Delta p = 0$ at the same point), this varies for the real profiles (e-h).
}
\end{figure*}

\begin{figure*}[h]
\centerline{
\includegraphics[width=0.5\textwidth,angle=0,clip]{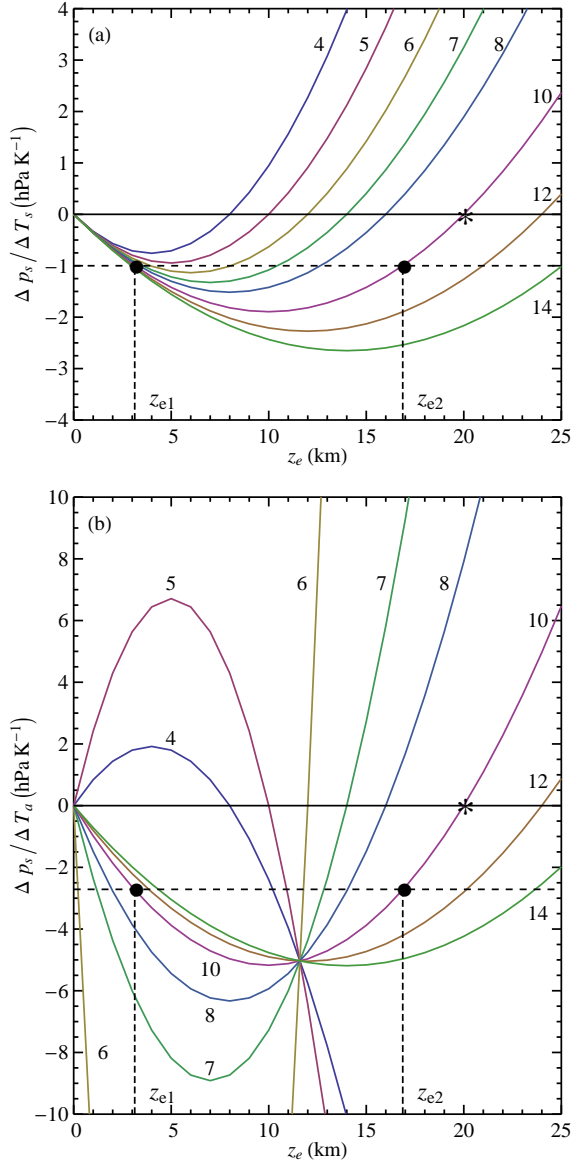}
}
\caption{\label{figteor}
Dependence of $\Delta p_s/\Delta T_s$ (a, Eq.~\ref{a/b}) and $\Delta p_s/\Delta T_a$ (b, Eqs.~\ref{a/n}, \ref{dy}) on isobaric height $z_e$ for different
values of isothermal height $z_i$ (km) that are shown near the corresponding curves. Vertical dashed lines denote isobaric heights $z_{e1}$ and $z_{e2}$
corresponding to the tropical mean $\Delta p_s/\Delta T_s = -1$~hPa~K$^{-1}$ (horizontal dashed line in (a), Fig.~\ref{figgra}d) for $z_i = 10$~km.
With an increase in $z_{e2}$ from 17~km to 20~km for $z_i = 10$~km both $\Delta p_s/\Delta T_s$ and $\Delta p_s/\Delta T_a$ become zero (shown as asterisk).
}
\end{figure*}

\begin{figure*}[h]
\centerline{
\includegraphics[width=1\textwidth,angle=0,clip]{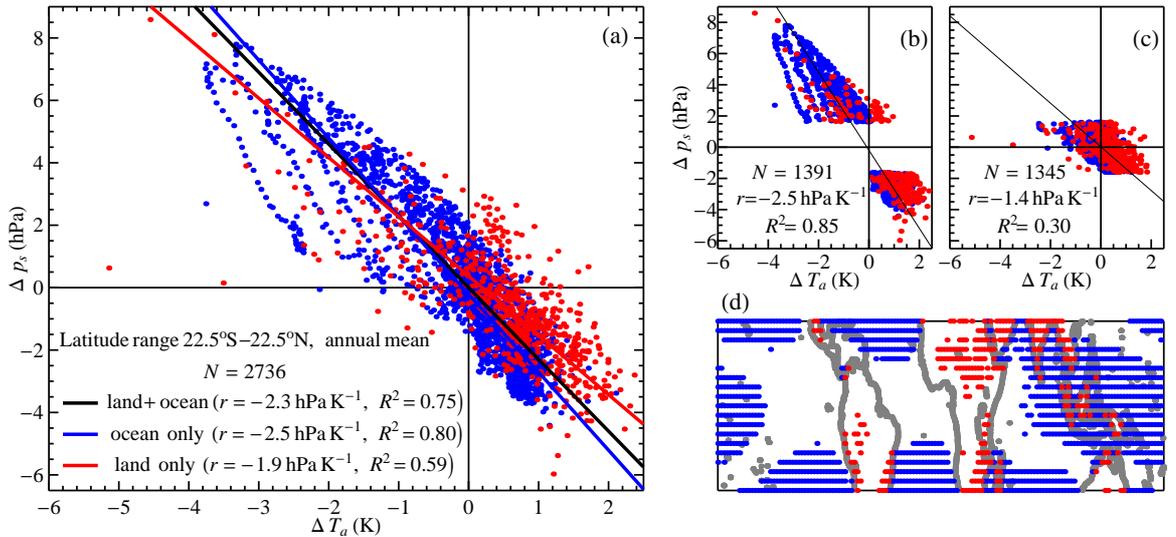}
}
\caption{\label{figplot}
Dependence of local sea level pressure $p_s$ on local mean tropospheric temperature $T_a$.
(a) Regression slopes $r$ and squared correlation coefficients $R^2$ for the linear regression $\Delta p_s = r \Delta T_a$,
where $\Delta p_s = p_s - \overline{p_s}$ and $\Delta T_a = T_a - \overline{T_a}$,
where all local values are annual means and the overbars denote averaging over the latitude range 22.5$^{\rm o}$S and 22.5$^{\rm o}$N,
for land (red), ocean (blue) and area as a whole (black).
(b): $r$ and $R^2$ for regression of data points where $|\Delta p_s|$ is greater than 1.6~hPa, which is equal to 2/3 standard deviation of the
frequency distribution of $\Delta p_s$ values in (a). (c): The same as (b) but for data points where $|\Delta p_s|$ is smaller than 1.6~hPa.
(d): Geographic location of the grid points from panel (b). Empty space corresponds to grid points from panel (c).
}
\end{figure*}

\begin{figure*}[h]
\centerline{
\includegraphics[width=1\textwidth,angle=0,clip]{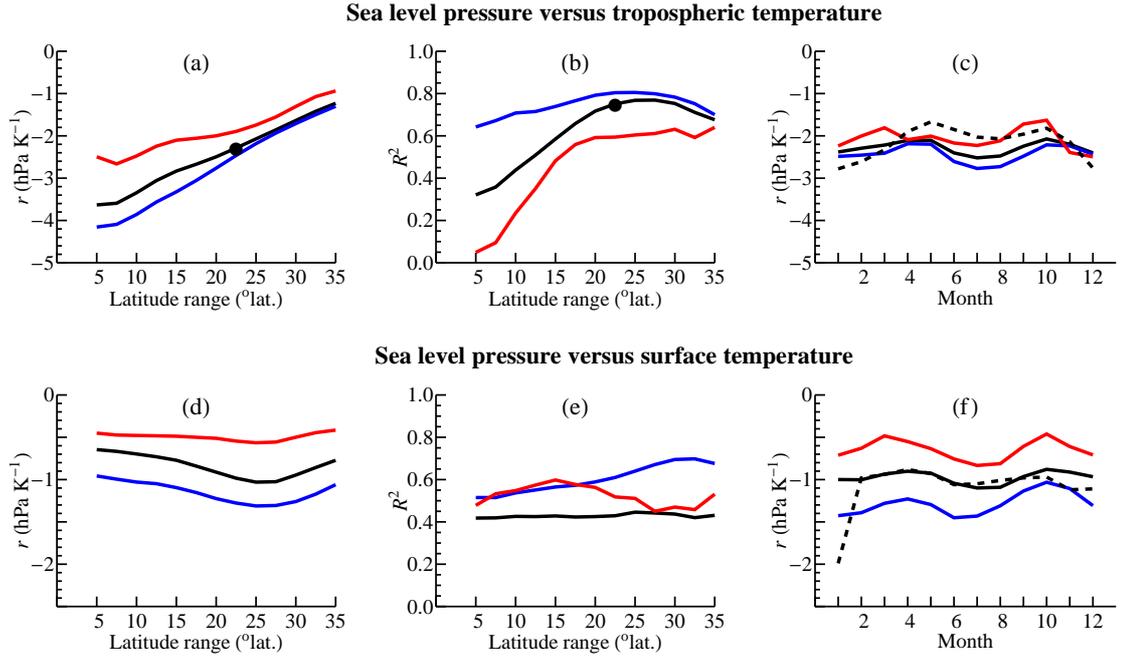}
}
\caption{\label{figgra}
(a,b) and (d,e): Regression slopes $r$ and $R^2$ for the regression $\Delta p_s = r \Delta T_a$ (a,b) and $\Delta p_s = r \Delta T_s$ (d,e) for
land (red), ocean (blue) and total area (black) as dependent on the considered latitude range.
E.g. latitude range 35\degree lat. means that averaging is made from  35$^{\rm o}$S to 35$^{\rm o}$N.
Here $\Delta X \equiv X - \overline{X}$ ($X = p_s,\,T_a,\,T_s$) where $X$ is the local annual mean value and $\overline{X}$ is
its spatial average over the considered latitude range.
Data corresponding to Fig.~3 of \citet{bayr13} (and to our Fig.~\ref{figplot}a) are marked with black circle. (c,f)
Seasonal variation of $r$ for the latitude range 22.5$^{\rm o}$S
and 22.5$^{\rm o}$N: land (red), ocean (blue) and total tropics (black) curves
denote results as in panels (a,d) but for particular months; the dashed line denotes the regression slope of zonally averaged
$\Delta p_s$ and $\Delta T_a$ ($\Delta T_s$).
}
\end{figure*}

\begin{figure*}[h]
\centerline{
\includegraphics[width=0.90\textwidth,angle=0,clip]{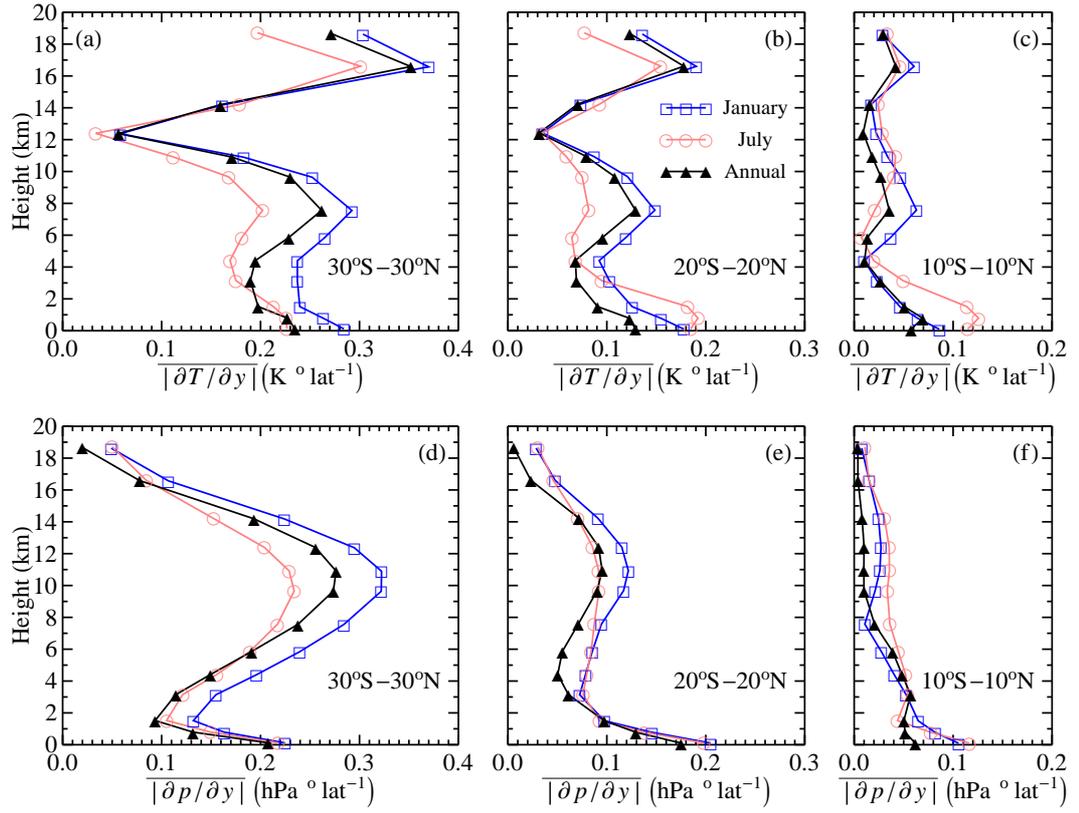}
}
\caption{\label{figtemp}
Vertical profiles of the meridional temperature (a-c) and pressure (d-f) gradients
taken by absolute magnitude and averaged from $30\degree$S to $30$$\degree$N (a,d),
$20\degree$S to $20\degree$N (b,e) and $10\degree$S to $10\degree$N (c,f) in January (blue squares), July (pink circles)
and annually (black triangles).
}
\end{figure*}

\begin{figure*}[h]
\centerline{
\includegraphics[width=1\textwidth,angle=0,clip]{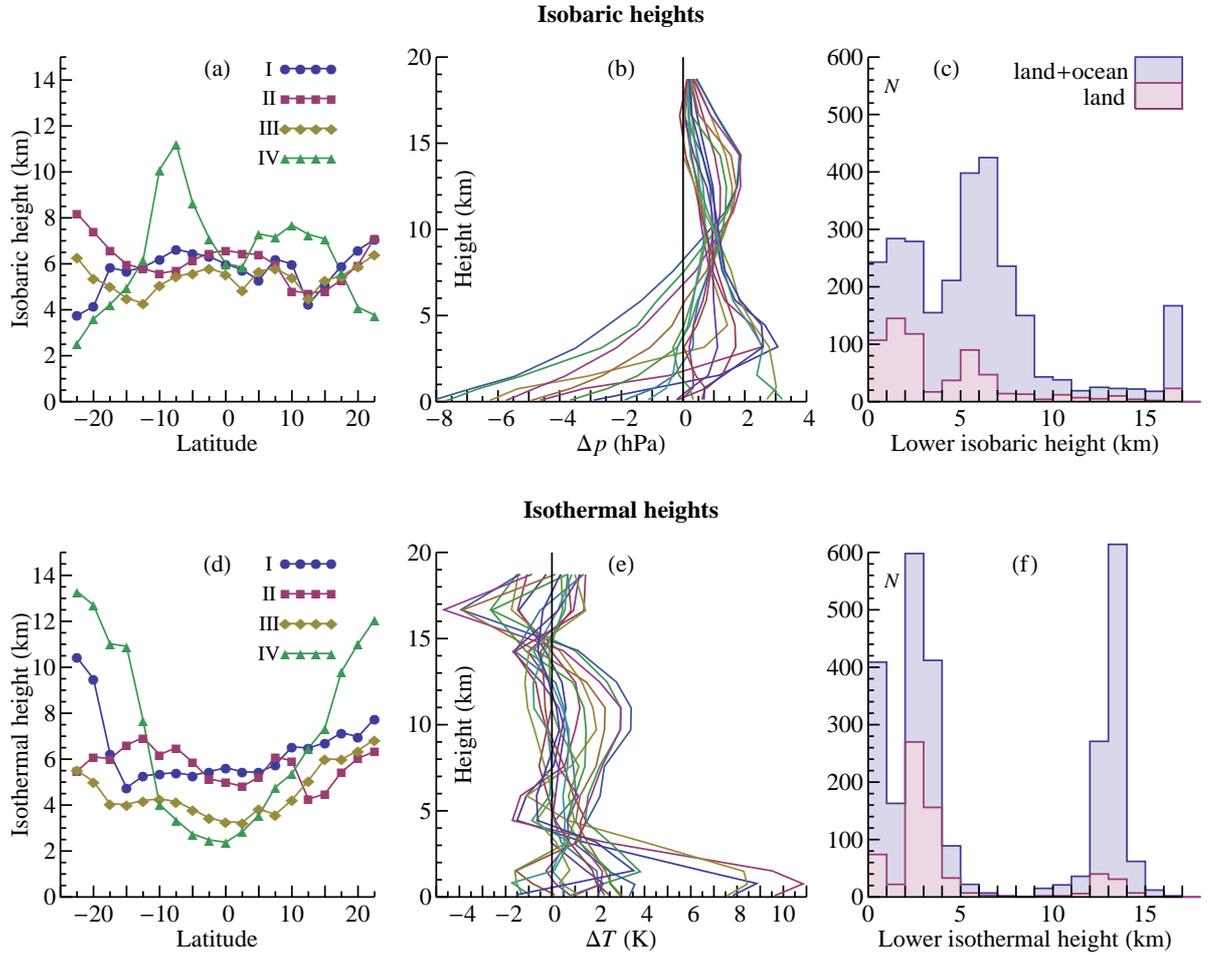}
}
\caption{\label{figiso}
Isobaric (a-c) and isothermal (d-f) heights in the tropical troposphere. (a),(d): zonal averages of isobaric and isothermal heights
defined as follows: I -- minimum height where local pressure (temperature) coincides with the mean pantropical pressure (temperature) at this height,
II -- minimum height where local pressure (temperature) coincides with the mean zonal pressure (temperature) at this height,
III -- minimum height where local zonal gradient of pressure (temperature) is zero, IV -- minimum height where local meridional gradient of pressure
(temperature) is zero. All local pressure and temperature values are annual means. (b), (e): vertical profiles of the
differences between the mean pantropical profile of pressure (temperature) in July and the profile
of pressure (temperature) for 18 individual grid points at 20$^{\rm o}$N that are spaced by 20$^{\rm o}$ longitude
starting from $0$\degree E.
Note the difference between isobaric heights I and II: individual profiles coincide with each other at a different height (around 10~km) than
they coincide with the pantropical mean profile. (c), (f): frequency distribution of isobaric and isothermal heights 1, the inner histogram shows
the distribution of land values only and the outer histogram shows all values.
}
\end{figure*}

\begin{figure*}[h]
\centerline{
\includegraphics[width=0.95\textwidth,angle=0,clip]{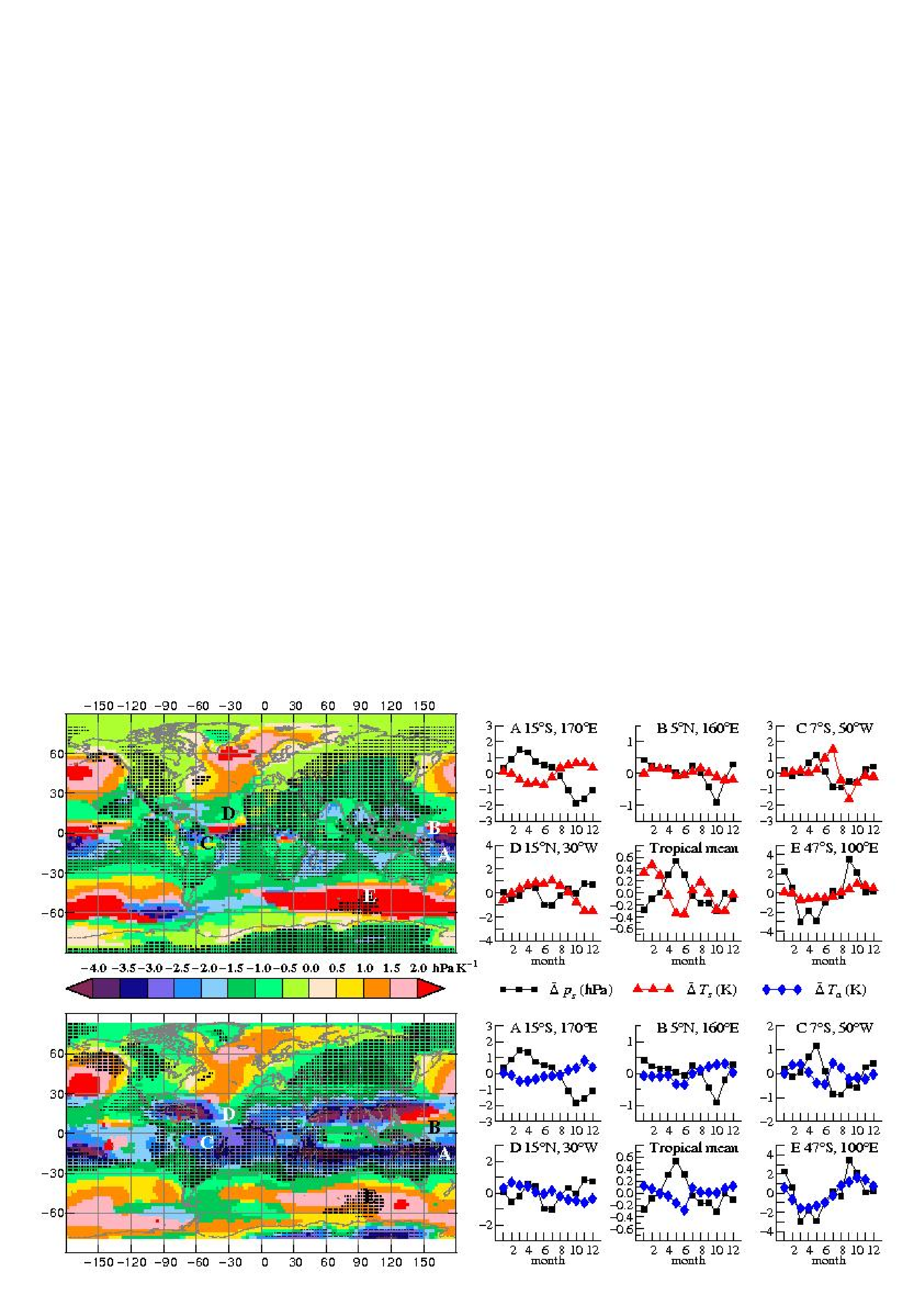}
}
\caption{\label{figmap}
Mean ratio between local monthly changes of sea level pressure $p_s$ and surface temperature $T_s$ (larger top left panel) and
$p_s$ and tropospheric temperature $T_a$ (larger lower left panel). The ratio is estimated as the slope coefficient of a Reduced Major
Axis regression of $\widetilde{\Delta} p_s$ on, respectively,
$\widetilde{\Delta}T_s$ and $\widetilde{\Delta} T_a$, see Section~\ref{tem} for details.
Black dots indicate where the probability level of the regression is less than $0.01$.
The small panels exemplify seasonal changes of $\widetilde{\Delta}p_s$, $\widetilde{\Delta}T_s$ and $\widetilde{\Delta}T_a$
in individual grid points (A, B, C, D and E) shown in the big panels, as well as the tropical mean (the area between 22.5$\degree$S
and 22.5$\degree$N). Note the different vertical scales in the small panels.
}
\end{figure*}

\begin{figure*}[h]
\centerline{
\includegraphics[width=0.5\textwidth,angle=0,clip]{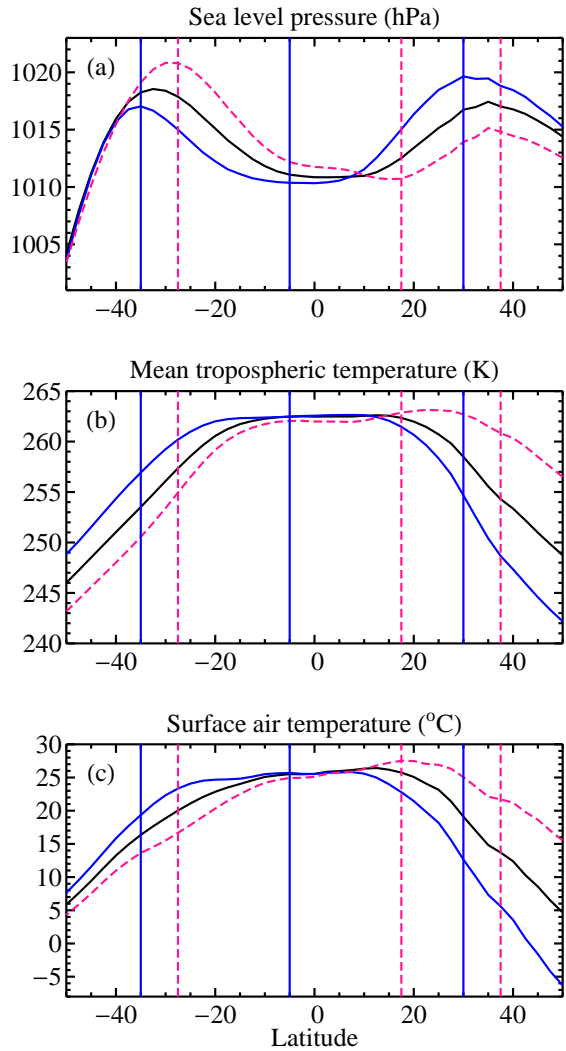}
}
\caption{\label{fighad}
Zonally averaged atmospheric parameters of Hadley cells. Solid black curve: annually averaged data,
solid blue curve: January, dashed pink curve: July. Vertical lines show the borders of the Southern
and Northern cells in January (solid blue) and July (dashed pink) defined as
the poleward maxima (the outer borders) and the central minimum (the inner border) of sea level pressure (a).
Monthly data from NCAR-NCEP reanalysis averaged for 1978-2013.
}
\end{figure*}

\begin{figure*}[h]
\centerline{
\includegraphics[width=0.9\textwidth,angle=0,clip]{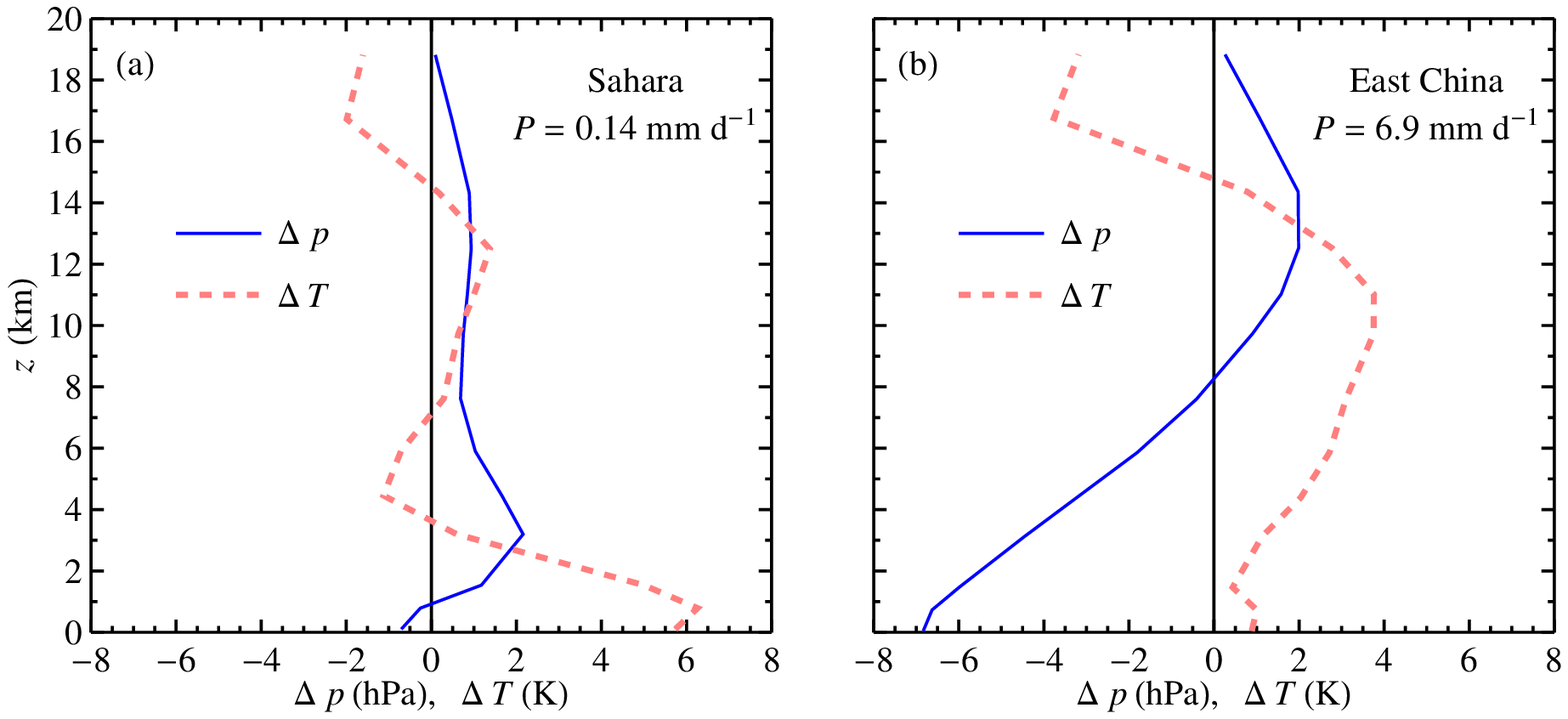}
}
\caption{\label{figsah}
Vertical profiles of air pressure and temperature differences between the zonal mean (20\degree N - 30\degree N)
in July and (a) Sahara (20\degree N - 30\degree N, 0\degree E - 20 \degree E) and (b) East China
(20\degree N - 30\degree N, 100\degree E - 120 \degree E). $P$ is precipitation in July in these regions.
}
\end{figure*}

Taking the derivative of (\ref{Ta2}) over $T_s$ and $c$ we obtain:
\beq\label{c}
dc = \frac{db-dn}{1 - Z/(e^Z-1)},\,\,\,dn \equiv \frac{dT_a}{T_a}.
\eeq
For the height of the tropical troposphere $z = H = 16.5$~km, $T_s = 298$~K and $\Gamma = 6.1$~K~km$^{-1}$ (Fig.~\ref{figlap})
we have $Z=1.9$, $c = 0.18$, $1 - Z/(e^Z-1) = 0.66$ and obtain from
(\ref{Ta2}) and (\ref{c})
\beq\label{Ta3}
T_a = 0.88 T_s,\,\,\,db = dn + 0.66 dc,\,\,\,\frac{dT_s}{dT_a} = \frac{1}{0.88}\left(1+0.66 \frac{d\Gamma}{dT_a}\frac{T_a}{\Gamma_g}\right).
\eeq
The mean tropospheric $\overline{T_a} = 262$~K in the tropics estimated from (\ref{Ta3}) is identical
to the annual tropical mean $\overline{T_a} = 262$~K (22.5\degree S -- 22.5\degree N) that we estimate from the TTT data of
\citet{mears09} and close to $\overline{T_a} = 263.6$~K cited by \citet{bayr13}.
\end{appendix}

\bibliography{met-refs}

\end{document}